\documentclass[sigconf]{acmart} %

\usepackage{multirow}
\usepackage{graphicx}
\usepackage{tikz}
\usetikzlibrary{shapes.geometric, arrows.meta, positioning}

\AtBeginDocument{%
  }

\setcopyright{acmlicensed}
\copyrightyear{2024}
\acmYear{2024}
\acmDOI{XXXXXXX.XXXXXXX}

\acmConference[CIKM '24]{3rd International Workshop on Industrial Recommendation Systems (at CIKM 2024)}{Oct 21--25, 2024}{Boise, Idaho, USA}

\acmISBN{978-1-4503-XXXX-X/18/06}

\begin{document}

\title{Improving Pinterest Search Relevance Using Large Language Models}

\author{Han Wang}
\authornotemark[1]
\email{hanwang@pinterest.com}
\affiliation{%
  \institution{Pinterest}
  \city{San Francisco}
  \country{USA}
}

\author{Mukuntha Narayanan S}
\authornote{These authors contributed equally to this research.}
\email{mukuntha@pinterest.com}
\affiliation{%
  \institution{Pinterest}
  \city{San Francisco}
  \country{USA}
  }

\author{Onur Gungor}
\email{ogungor@pinterest.com}
\affiliation{%
  \institution{Pinterest}
  \city{San Francisco}
  \country{USA}
}

\author{Yu Xu}
\email{yuxu@pinterest.com}
\affiliation{%
  \institution{Pinterest}
  \city{San Francisco}
  \country{USA}
}

\author{Krishna Kamath}
\email{kkamath@pinterest.com}
\affiliation{%
  \institution{Pinterest}
  \city{San Francisco}
  \country{USA}
}

\author{Rakesh Chalasani}
\email{rchalasani@pinterest.com}
\affiliation{%
  \institution{Pinterest}
  \city{San Francisco}
  \country{USA}
}

\author{Kurchi Subhra Hazra}
\email{ksubhrahazra@pinterest.com}
\affiliation{%
  \institution{Pinterest}
  \city{San Francisco}
  \country{USA}
}

\author{Jinfeng Rao}
\email{marquisrao@pinterest.com}
\affiliation{%
  \institution{Pinterest}
  \city{San Francisco}
  \country{USA}
}

\renewcommand{\shortauthors}{Wang and Narayanan, et al.}

\begin{abstract}

To improve relevance scoring on Pinterest Search, we integrate Large Language Models (LLMs) into our search relevance model, leveraging carefully designed text representations to predict the relevance of Pins effectively. Our approach uses search queries alongside content representations that include captions extracted from a generative visual language model. These are further enriched with link-based text data, historically high-quality engaged queries, user-curated boards, Pin titles and Pin descriptions, creating robust models for predicting search relevance. We use a semi-supervised learning approach to efficiently scale up the amount of training data, expanding beyond the expensive human labeled data available. By utilizing multilingual LLMs, our system extends training data to include unseen languages and domains, despite initial data and annotator expertise being confined to English. 
Furthermore, we distill from the LLM-based model into real-time servable model architectures and features. We provide comprehensive offline experimental validation for our proposed techniques and demonstrate the gains achieved through the final deployed system at scale.

\end{abstract}




\maketitle

\section{Introduction}

Search relevance measures how well the search results align with the intent behind the search query. Using a relevance objective allows search engines to ensure that the content displayed to users is genuinely pertinent to their information needs. Without relevance scores, search engines might overly rely on factors like past user engagement, leading to results skewed towards click-worthy or sensational content rather than truly relevant information, compromising the quality and usefulness of a search engine.

Pinterest Search is one of the key surfaces on Pinterest where users can discover inspiring contents that align with their information needs. Delivering a relevant search feed helps better fulfill users' intent and bring them the inspiration to create a life they love. The visual discovery nature of Pinterest Search poses unique challenges, as most content on the platform is present in the format of images or videos. Additionally, Pinterest Search serves a global audience in real-time, needing to accommodate users who speak over 45 different languages with diverse cultural backgrounds and interests using Pinterest for visual discovery.

To measure the relevance between queries and Pins, we use a 5-level guideline, where higher levels indicate better relevance. Compared to binary relevance judgements, such fine-grained relevance judgments can better capture the complex relationship between queries and Pins. Based on this guideline, we build a search relevance model by fine-tuning Large Language Models (LLMs) to predict the relevance scores given a query and the text representation of a Pin. We incorporate carefully designed text representation for each Pin, including Pin titles, descriptions, image captions generated by a generative visual language model, link-based text data, user-curated boards, and historically high-quality engaged queries. These enriched text features contribute to a more robust relevance model. 
In real-time serving, however, the powerful LLMs come with high latencies and computational costs. Therefore, we apply the knowledge distillation technique where the relevance scores are distilled from the LLM-based teacher model into a real-time servable student model. The LLM-based teacher model can greatly scale up the training data and also expand to languages and domains that were not initially covered in the human-annotated data, which greatly enhances the performance of the production relevance model.

We would like to highlight our contributions as follows:
\begin{itemize}
    \item We build an LLM-based relevance system for Pinterest Search and validate its performance and effectiveness through extensive offline experiments and online A/B tests.
    \item We exploit the potential of incorporating enriched text features and metadata to build a robust search relevance model.
    \item We empirically demonstrate the benefit of knowledge distillation in scaling up training data and generalizing the data to various languages and countries.
\end{itemize}

\section{Related Work}

Relevance modeling can be categorized into token-based and neural model-based approaches. Token-based methods, like TF-IDF and BM25 \cite{robertson1995okapi,robertson2009probabilistic}, focus on term matching but lack semantic understanding and often fail to capture the full context of the search. Neural model-based approaches leverage neural networks and language models to better capture nuanced semantic relationships between queries and documents, which have made significant progress in relevance modeling \cite{zhu2023large}. 

In language model-based relevance modeling, there are mainly two types of approaches used in practice: representation-based and interaction-based. Representation-based models, also known as bi-encoder models, independently encode a query and document into a common dense space and then score their relevance using vector dot-product or cosine similarity. These models are typically trained by minimizing the contrastive loss with in-batch negative sampling \cite{karpukhin2020dense,ma2023fine}. Although representation-based models are efficient for retrieval tasks, they have limited representational power to capture the intricate interactions between queries and documents. 

Interaction-based models, also known as cross-encoder models, jointly encode queries and documents. Compared to bi-encoder models, these models can better capture the interaction between queries and documents, hence are commonly used in the re-ranking stage. For instance, monoBERT \cite{nogueira2019passage} uses BERT to encode concatenated query-document pairs and fine-tunes the model to minimize pointwise binary classification loss. monoT5 \cite{nogueira2020document} builds on this by using the sequence-to-sequence T5 model \cite{2020t5} to directly output relevance labels based on query-document inputs, and RankT5 \cite{zhuang2023rankt5} further optimizes ranking performance using pairwise and listwise ranking loss. RankLLaMA \cite{ma2023fine} fine-tunes the LLaMa model with contrastive loss, achieving state-of-the-art performance on several public benchmark datasets. In this work, we adopt a cross-encoder architecture and fine-tune the LLMs by minimizing multi-class classification loss. 

Cross-encoder models with LLM backbones have demonstrated significant improvements in relevance modeling performance. However, their high computational cost and latency during inference limit their deployment in real-world search scenarios. To enable real-time serving, one strategy is to apply knowledge distillation, transferring knowledge from a large teacher model to a smaller student model and augmenting existing training data with teacher-generated labels \cite{tang2019distilling,hofstatter2020improving,gao2020understanding,reddi2021rankdistil}. 
In our work, we distill from an LLM-based cross-encoder teacher model to a lightweight student model. The student model incorporates several in-house query embeddings and Pin embeddings, as well as text-matching features like BM25 match scores. This approach greatly reduces online serving latency while being able to narrow the performance gap with the teacher model by scaling up the augmented teacher labels.

\section{Methodology}

\subsection{Problem Statement}

In our work, we formulate search relevance prediction as a multi-class classification problem. The model categorizes Pins in response to a search query into five ordered relevance levels: Excellent / Highly Relevant (L5), Good / Relevant (L4), Complementary / Marginally Relevant (L3), Poor / Irrelevant (L2), and Highly Irrelevant (L1), as detailed in Table \ref{tab:pin_relevance}. Training and evaluation labels are obtained through human annotators.

\begin{table*}
  \caption{Five-scale Pin relevance guidelines.}
  \label{tab:pin_relevance}
  \scalebox{0.93}{\begin{tabular}{cl}
    \toprule
    Relevance Label & Description \\
    \midrule
    Excellent / Highly Relevant (L5) & Exactly matches or directly associates with the search query. \\
    Good / Relevant (L4) & Close match or a potential substitute to the search query, with slight mismatches. \\
    Complementary / Marginally Relevant (L3) & Related to the search query but only partially matches it, not specifically addressing the intent. \\
    Poor / Irrelevant (L2) & Fitting into the general category but not serving the intended purpose or matching the user intent. \\
    Highly Irrelevant (L1) & Completely irrelevant to the search, potentially causing user dissatisfaction  \\
  \bottomrule
\end{tabular}}
\end{table*}

\subsection{LLM as Relevance Model}
\subsubsection{Model Architecture}
\label{sec:model_architecture}

\begin{figure}[ht]
  \centering
  \includegraphics[width=0.4\linewidth]{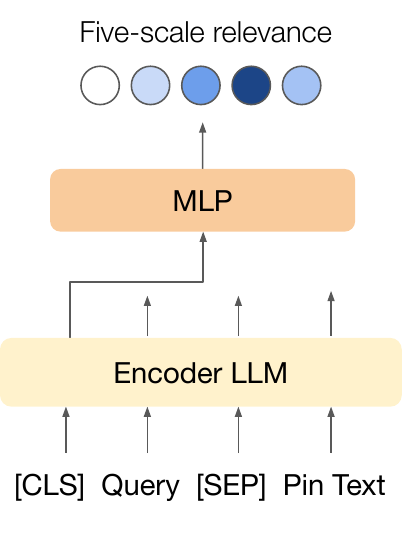}
  \caption{The cross-encoder architecture in the relevance teacher model. Take the encoder language models (e.g., BERT-based models) for illustration.}
  \label{fig:model_architecture}
\end{figure}

We leverage language models to predict the relevance of a Pin to a search query based on textual information. To achieve this, we utilize a cross-encoder structure \cite{nogueira2019passage} that encodes the query and Pin text together, as depicted in Figure \ref{fig:model_architecture}. Specifically, we concatenate the query text and Pin text with a separator token. The tokenized text is then fed into the language model to generate a sentence embedding. For encoder-based language models, we use the embedding of the [CLS] token (or <s>) as the sentence embedding. In the case of decoder-based language models, we use the embedding of the final non-padding token. This embedding is then passed through several fully-connected layers. The output dimension corresponds to the five relevance levels, and we apply softmax to obtain the relevance scores. 
During training, we fine-tune the pre-trained language models by minimizing pointwise multi-class cross-entropy loss.

\subsubsection{Pin Text Representations}
\label{sec:pin_text_signals}

Pins on Pinterest are rich multimedia entities that feature images, videos and other content, often linked to external webpages or blogs. Accompanying the image, each Pin also includes a title, a description, and even user-generated comments. To represent each Pin, we use the following varied set of text features derived from metadata, the image itself, as well as user-curated data. These features are designed with a focus on providing reliable high-quality representations, while retaining high coverage across Pins on Pinterest Search.

\begin{enumerate}
    \item \textbf{Pin titles and descriptions}: These are the title and the description for the Pin set by the user who created the Pin.
    \item \textbf{Synthetic image captions}: We use an image captioning model to generate synthetic image descriptions for the images. In our experiments, we use captions generated by BLIP \cite{li2022blip}, an off-the-shelf image captioning model. An evaluation by human assessors on a sample of 10,000 images showed that 88\% of these captions were both high-quality and relevant to the image.
    \item \textbf{High-engagement query tokens}: We take the last 2 years' worth of queries that had the highest engagements with this Pin on the search surface, and collect all the unique tokens across all of them. To maximize the quality of these representations, we only use tokens from the queries with the most engagements, and consider high-quality engagements such as repins (when a user saves a Pin to a board) and long clicks (where users clicked into the linked page for more than 10 sec).
    \item \textbf{User-curated board titles}: On Pinterest, users curate their own personal collections called Boards, saving Pins to them and giving them a title. In the past, we have demonstrated the usefulness of the interactions shown by this human-curated content \cite{ying2018graph}, and in this work, we leverage the titles for each board that a Pin has saved to. 
    \item \textbf{Link titles and descriptions}: A key metric optimized for Pinterest Search is the "long click", which occurs when a user clicks through to a Pin's linked webpage and spends over ten seconds there. This behavior has historically correlated strongly with search relevance, underlining the importance of linked content in relevance modeling. Consequently, we integrate the titles and descriptions of URLs into our models. Our experiments found that these link titles and descriptions are particularly effective in automatically imputing missing titles and descriptions on Pins.
\end{enumerate}

\subsection{Distill LLM into Student Model}
\label{subsec:student_model}

\begin{figure}[htb]
  \centering
  \includegraphics[width=0.95\linewidth]{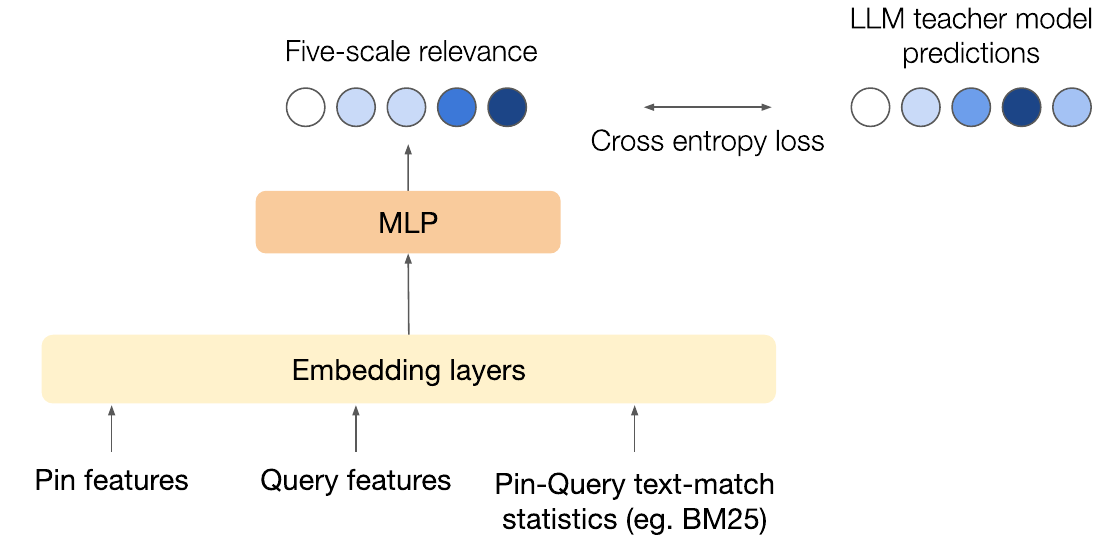}
  \caption{The online-served student model is trained via distillation from the LLM-based relevance teacher model.}
  \label{fig:student_model_architecture}
\end{figure}

Since our cross-encoder language model based classifier is hard to scale to Pinterest Search scale given real time latency and cost considerations, we use a teacher-student distillation setup to distill into a student relevance model, as shown in Figure \ref{fig:student_model_architecture}.

\subsubsection{Model Architecture} 
\label{subsec:student_model_architecture}

The student model is based on a feed-forward neural network that includes online ranking features such as query embeddings and Pin embeddings, as well as numerical and categorical features (described in Section \ref{subsec:online_serving}). We transform our numerical features into embeddings using linear transforms \cite{gorishniy2021revisiting} and also learn embeddings for any categorical features.

\subsubsection{Knowledge Distillation and Semi-Supervised Learning}
\label{subsec:knowledge_distillation}

To train our online relevance model, described in Section \ref{subsec:student_model_architecture}, we utilize logged data from search engagements and impressions. This vast dataset with billions of rows lacks relevance labels but includes several real-time logged features for the search request and the Pin. 

Here, we employ the LLM-based teacher model to generate soft 5-scale relevance labels, thereby augmenting our training dataset to over 100 times the size of our human-labeled data pool. 
We also employ a stratified sampling strategy to ensure a balanced representation across all relevance labels in our training set. 
The final training dataset contains the search query and a Pin along with the relevance predictions from the teacher model. 

\begin{figure*}[ht]
  \centering
  \includegraphics[width=0.6\linewidth]{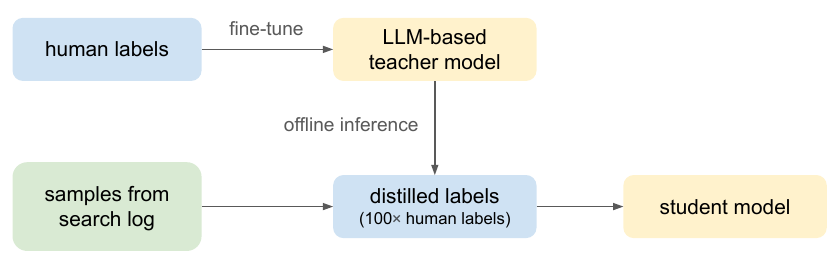}
  \caption{Overview of the proposed search relevance system at Pinterest.}
  \label{fig:knowledge_distill}
\end{figure*}

The labeled and balanced dataset is subsequently used to train a smaller, more computationally efficient student model optimized for online serving. This student model is crucial for its ability to predict relevance scores with low latency using real-time features. An overview of the search relevance system is illustrated in Figure \ref{fig:knowledge_distill}. The relevance scores generated by this model are then utilized alongside engagement predictions to determine the final ranking of search results. 

This blend of knowledge distillation and semi-supervised learning not only makes effective use of vast amounts of initially unlabeled data. The user logged data, unlike human-labeled relevance data, also includes a wide range of languages from around the world and new concepts not encountered in our human-labeled data owing to the seasonality in Pinterest Search. By using a multilingual LLM-based teacher model, we are able to successfully generalize from human-labeled data focused on US queries to unseen languages and countries.

\section{Experiments}

\subsection{Experiment Setup}

\subsubsection{Datasets}

The search relevance teacher model is trained and evaluated using human-annotated labels following the 5-scale relevance guidelines. Given the accuracy of a single rater is only around 70\%, we train the model using soft labels aggregated from 3 raters. Additionally, we apply a stratified sampling strategy based on annotated labels to obtain a more balanced dataset. 
The data is then randomly split into a training set and a test set based on queries, allowing us to better assess the model's performance on unseen queries. While several publicly available datasets exist for relevance modeling \cite{nguyen2016ms,chen2022wands,reddy2022shopping}, they do not contain those Pinterest-specific text features. Therefore, we will focus on in-house data for the rest of the discussions. Table \ref{tab:human_label_stats} presents the label distribution in this human-annotated dataset.

\begin{table}[ht]
  \caption{Statistics of the human-annotated relevance dataset.}
  \label{tab:human_label_stats}
  \scalebox{1.00}{\begin{tabular}{ccccccc}
    \toprule
    Dataset & Total & L1 & L2 & L3 & L4 & L5 \\
    \midrule
    Train & 280,934 & 16.0\% & 19.7\% & 28.9\% & 13.1\% & 22.3\% \\
    Test & 31,551 & 15.9\% & 20.0\% & 28.9\% & 13.1\% & 22.1\% \\
  \bottomrule
\end{tabular}}
\end{table}

\subsubsection{Model Implementation} 

We use an LLM-based cross-encoder model as the relevance teacher model, with the model architecture described in Section \ref{sec:model_architecture}. In this experiment, we evaluate the following pre-trained language models: multilingual BERT$_{base}$ \cite{devlin2018bert}, T5$_{base}$ \cite{2020t5}, mDeBERTaV3$_{base}$ \cite{he2021debertav3}, XLM-RoBERTa$_{large}$ \cite{conneau2019unsupervised}, and Llama-3-8B \footnote{\url{https://huggingface.co/meta-llama/Meta-Llama-3-8B}}. These models are initialized from Hugging Face checkpoints and fine-tuned using our in-house search relevance training data. The training is conducted on 8 × 40G A100 GPUs. We utilize the FusedAdam optimizer \footnote{\url{https://nvidia.github.io/apex/optimizers.html\#apex.optimizers.FusedAdam}} and apply distributed data parallel (DDP) for efficient model training. For larger language models such as Llama, we first load quantized model weights and then apply qLoRA \cite{dettmers2024qlora} for fine-tuning. Additionally, we incorporate gradient checkpointing \cite{chen2016training} and mixed precision techniques to further improve training efficiency and memory usage.

\subsubsection{Online Serving}
\label{subsec:online_serving}

The student model served online uses a range of query-level features, including query interest ontology features, shopping interest features, and SearchSAGE query embeddings \cite{agarwal2024omnisearchsage}. For Pin-level features, it uses PinSAGE embedding \cite{ying2018graph} (trained representations for each Pin using the GraphSAGE \cite{hamilton2017inductive} algorithm on the Pinterest board graph), visual embeddings for the image \cite{zhai2019learning,beal2022billion}, and SearchSAGE Pin embeddings \cite{agarwal2024omnisearchsage}. Additionally, the model incorporates several query-Pin interaction features, such as BM25 scores for different text fields, historical engagement rates between the Pin and the query, and the fraction of overlapping tokens between the query and different text fields. These features are embedded and passed through a feed-forward network to predict five-scale relevance scores. This model is currently served real-time. %

To train this model, we employ the stratified sampling strategy mentioned in Section \ref{subsec:knowledge_distillation} to produce a training dataset comprising 30M rows of teacher-labeled query-Pin pairs. Since the dataset is based on sampled user interaction data from around the world, it contains queries in more than 45 languages, spanning different countries, interests and cultural contexts.

\subsection{Experiment Results}

\subsubsection{Evaluation Metrics}

For all offline experiments, we report the accuracy of 5-scale relevance predictions as well as the AUROC metrics for binarized labels with thresholds at 3, 4, and 5, since correctly identifying highly relevant content is more important for search ranking. 

\subsubsection{Comparison of Language Models} 

Table \ref{tab:teacher_benchmark} shows the performance of different language models. As a baseline, we include a model that relies solely on the SearchSAGE embeddings \cite{agarwal2024omnisearchsage}. In this comparison, we keep the text features for each Pin and the maximum text length fixed, varying only the language models. The results in Table \ref{tab:teacher_benchmark} clearly demonstrate that the language models offer additional improvements over our in-house content and query embedding. Furthermore, more sophisticated language models and larger model sizes consistently enhance the relevance prediction performance. Specifically, the 8B LLaMa-3 model outperforms the multilingual BERT$_{base}$ model by 12.5\% and the baseline model by 19.7\% in terms of 5-scale accuracy.

\begin{table}[ht]
  \caption{Comparisons of different language models on 5-scale relevance prediction.  The AUROC metrics are reported for binarized labels with thresholds 3, 4, and 5. }
  \label{tab:teacher_benchmark}
  \scalebox{1.00}{
  \begin{tabular}{lcc}
    \toprule
    Model & Accuracy & AUROC 3+/4+/5+ \\
    \midrule
    \midrule
SearchSAGE & 0.503 & 0.878/0.845/0.826 \\  
\midrule
mBERT$_{base}$ & 0.535 & 0.887/0.864/0.861\\  
\midrule
T5$_{base}$ & 0.569 & 0.909/0.884/0.886 \\
\midrule
mDeBERTaV3$_{base}$ & 0.580 & 0.917/0.892/0.895 \\
\midrule
XLM-RoBERTa$_{large}$ & 0.588 & 0.919/0.897/0.900 \\
\midrule
Llama-3-8B & \textbf{0.602} & \textbf{0.930/0.904/0.908}\\
\bottomrule
\end{tabular}}
\end{table}

\subsubsection{Importance of Enriching Text Features}

To predict the relevance of a Pin to a query using only textual information, we enrich the Pin text representations with several carefully designed text features, detailed in Section \ref{sec:pin_text_signals}. 
We conduct an analysis to assess the impact of each text feature on relevance prediction, using mDeBERTaV3$_{base}$ as the language model and setting the maximum text length to 256. The results, summarized in Table \ref{tab:text_signal_ablation}, demonstrate that the model's performance consistently improves with the sequential addition of these text features. This indicates that enriched text features and metadata significantly contribute to building a more robust relevance model.

\begin{table}[ht]
  \caption{Benchmark the improvement with the sequential addition of text features.}
  \label{tab:text_signal_ablation}
  \scalebox{0.95}{\begin{tabular}{lcc}
    \toprule
    Text Features &  Accuracy & AUROC 3+/4+/5+ \\
    \midrule
    \midrule
    {Synthetic image caption} & 0.457 & 0.838/0.781/0.760 \\
    \midrule
    {+ Pin title and description}  & 0.561 & 0.906/0.875/0.876 \\
    \midrule
    {+ Link title and description} & 0.565 & 0.910/0.880/0.880 \\
    \midrule
    {+ User-curated board titles} & 0.577 & 0.916/0.888/0.890 \\
    \midrule
     {+ High-engagement query tokens} & 0.580 & 0.917/0.892/0.895 \\
  \bottomrule
\end{tabular}}
\end{table}

\subsubsection{Scaling Up Training Labels through Distillation} 

By using knowledge distillation and semi-supervised learning, as described in Section \ref{subsec:knowledge_distillation}, we can effectively scale the training data beyond the limited human-annotated data. The multilingual LLM-based teacher model is capable of generating training data for out-of-domain samples from different languages and countries, even though the annotated data is confined to English. To evaluate the effectiveness of this approach, we use a student model trained on human-annotated data as a baseline and experiment with varying sizes of distilled labels for training. We evaluate the models using the same test dataset as the teacher model. The results, summarized in Table \ref{tab:knowledge_distillation_comparsion}, demonstrate that training on augmented teacher-generated labels yields better performance compared to training solely on human-annotated data. Moreover, as the size of the training data increases, the student model consistently shows improved performance.

\begin{table}[ht]
  \caption{Comparisons of production model performance when training on different amounts of labels. }
  \label{tab:knowledge_distillation_comparsion}
  \scalebox{1.0}{\begin{tabular}{lcc}
    \toprule
    Training Data & Accuracy & AUROC 3+/4+/5+ \\
    \midrule
    \midrule
    0.3M human labels & 0.484 & 0.850/0.817/0.794 \\
    \midrule
    6M distilled labels & 0.535 & 0.897/0.850/0.841 \\
    \midrule
    12M distilled labels & 0.539 & 0.903/0.856/0.847 \\
    \midrule
    30M distilled labels & 0.548 & 0.908/0.860/0.850 \\
  \bottomrule
\end{tabular}}
\end{table}

\subsection{Online A/B Testing}

In this section, we present the results from A/B tests conducted, replacing the existing search relevance systems with the model trained using our newly proposed pipeline.

\subsubsection{Human Relevance Evaluations}

To understand the effectiveness of our new relevance model, we set up evaluations with human annotators to assess the relevance of the search feeds with and without the new relevance model serving traffic. For this evaluation, we selected a random set of 300 queries across both high-frequency (head) and low-frequency (tail) queries, using real user queries from a given country. We computed ranked lists for these queries with and without our model improvements, and measured the increase in normalized Discounted Cumulative Gain (nDCG) observed with our model. The results were evaluated at depths of 
20. We mapped the five relevance labels (L1 to L5) to 0, 0.25, 0.5, 0.75 and 1.0 respectively (see Table \ref{tab:pin_relevance}). In the nDCG computation, we normalized the DCG using an ideal DCG metric that assumes an infinite number of L5 relevant inventory available. 
Specifically, the nDCG@K here is calculated as follows
$$
nDCG@K=\frac{\sum_{k=1}^K 0.25(L - 1) / log_2(1+k)}{\sum_{k=1}^K 1 / log_2(1+k)}, \text{ }L\in\{1,2,3,4,5\}.
$$
Our proposed relevance modeling pipeline leads to a +2.18\% improvement in search feed relevance, as measured by nDCG@20.

In Table \ref{tab:judy_evals_country}, we report results of our Pin search feeds generated for users in different countries, as well as the improvements observed for queries with high shopping intent. Here we report the increases in 
precision@8
using the mapped five-scale relevance labels (0, 0.25, 0.5, 0.75, 1.0).
We use precision and a lower evaluation depth for these evaluations to control the evaluation costs.  
The results indicate that the multilingual LLM-based relevance teacher model effectively generalizes across languages not encountered during training. 

\begin{table}[ht]
  \caption{Human relevance judgements with search feeds seen across different countries demonstrate generalization across unseen languages.}
  \label{tab:judy_evals_country}
\begin{tabular}{@{}lll@{}}  
\toprule  
Segment                        & precision@8 \\ 
\midrule  
US & +1.5\% \\  
DE       & +0.64\% \\  
FR        & +0.84\%  \\  
UK        & +1.3\% \\  
\midrule
US: Shopping Interest Queries & +1.39\% \\  
\bottomrule
\end{tabular}  
\end{table}

\subsubsection{User Triggered Experiments}

In the context of Search experiments, the primary metric we focus on, in addition to relevance, is the search fulfillment rate. This metric is defined as the number of search sessions that result in a high-significance user action. 
In Table \ref{tab:fulfillment_rates}, we show that the improved relevance for the feed also results in big overall increases in search engagement and fulfillment. We break down results by country to show that the improvements we see in relevance also apply to non-US countries, despite not having annotated data available for those countries during model training.

\begin{table}[ht]
  \caption{Search Fulfillment Rate increases with the new relevance system show a significant uptick globally.}
  \label{tab:fulfillment_rates}
    \begin{tabular}{lc}  
        \toprule  
        Segment & Fulfillment Rate (A/B) \\  
        \midrule  
        US Traffic    & +0.7\% \\  
        Non-US Traffic & +2.0\% \\  
        \bottomrule  
    \end{tabular}  
\end{table}

\section{Conclusion}

In this work, we present an LLM-based relevance system for Pinterest Search. We thoroughly describe each building block of this system, including training data collection, model architecture, enriched text features, augmented label generation, and online serving. We conduct extensive offline experiments to validate the effectiveness of each modeling decision. Lastly, we present the results from online A/B experiment, which shows an improvement of >1\% in search feed relevance and >1.5\% in search fulfillment rates. To further enhance the efficacy of our relevance system, future work will explore the integration of serveable LLMs, vision-and-language multimodal models (VLLMs), and active learning strategies to dynamically scale and improve the quality of the training data.

\bibliographystyle{ACM-Reference-Format}
\bibliography{reference}

\appendix

\end{document}